\documentclass[aps,twocolumn,showpacs,floatfix]{revtex4}

  \usepackage{graphics,graphicx,epsfig}

\begin{document}

%\twocolumn[
\title{Coherent analysis of quantum optical sideband modes}
\author{E. H. Huntington, G. N. Milford, C. Robilliard}
\affiliation{Centre for Quantum Computer Technology, School of
Information Technology and Electrical Engineering, University
College, The University of New South Wales, Canberra, ACT, 2600}
\author{T. C. Ralph}
\affiliation{Centre for Quantum Computer Technology,  Department of
Physics, University of Queensland,  St Lucia QLD 4072 Australia}

\begin{abstract}
 We demonstrate a device that allows for the coherent analysis of a pair of optical frequency
 sidebands in an arbitrary basis.  We show that our device is quantum noise limited and hence applications for this scheme may be
 found in discrete and continuous variable optical quantum information experiments.\\
	
\end{abstract}

%\ocis{270.0270, 270.5570}

\maketitle
%]

%\newpage

%\section{Introduction}

Quantum information can be encoded on light using two orthogonal optical modes \cite{KLM}. Polarization is a good example and can be used to encode both discrete quantum bits (qubits) \cite{jam} and continuous variable quantum information \cite{koro,bow}. A key feature of polarization is the ability to analyse in any basis using a series of half and quarter-wave plates (or a rotatable Babinet compensator), a polarizing beamsplitter and intensity measurements. By analysing in a complete set of bases, say linear, diagonal and circular, the amplitudes and relative phase of the polarization modes can be determined. Another example encoding scheme is a pair of optical sideband modes separated from an average or carrier frequency by a radio or microwave frequency.  Much like polarisation analysis, full characterisation of this encoding scheme would reveal the amplitude and relative phase of the sidebands. However unless here is equal power in the two frequency modes, homodyne detection of the in-phase and out-of-phase quadratures is insufficient to fully characterize the system.

In this work we propose and demonstrate a device designed to achieve full characterization of optical sideband modes. This "rf-analyser" may be used to analyse the average optical power in each optical sideband mode and the phase relationship between them.   The rf-analyser is directly analogous to a polarisation analysis system comprising a rotatable Babinet compensator followed by a polarising beamsplitter \cite{Hecht}. We show that our device is quantum noise limited and thus could be used as a new analysis tool for non-classical states of light both in the continuous variable and the discrete variable \cite{hunt04} domains.

%\section{Theory}

\begin{figure}[htb]
\centerline{ \includegraphics[width=8.3cm]{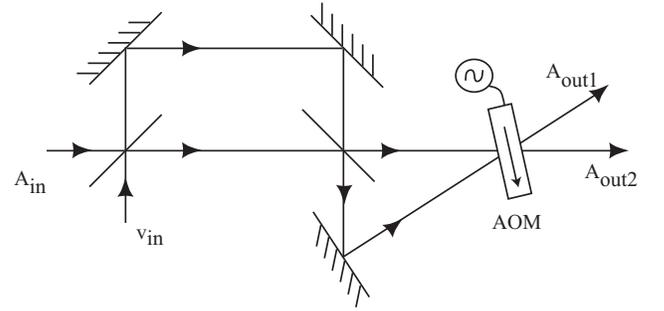}}
  \caption
{ \label{schem} A schematic diagram of the system.  All parameters are defined in the text.}
 \end{figure}

\noindent {\it Theory}: Fig. \ref{schem} schematically illustrates the system.  It is essentially an unequal arm-length Mach-Zehnder interferometer (UMZI) followed by an acousto-optic modulator (AOM) with the two UMZI outputs interferometrically combined at the AOM.  In the Heisenberg picture this system has two inputs: $\hat{A}_{in}$ and $\hat{v}_{in}$; and two outputs: $\hat{A}_{out1}$ and $\hat{A}_{out2}$ .  Let us define the average or carrier frequency to be  $\omega_0$ and the radio or microwave sideband frequencies of interest to be $\pm \Omega$.    The annihilation operators for the outputs of the rf-analyser at the frequencies of interest are 

\begin{eqnarray}
	\label{out} 
	A_{out1}(\Omega) &=& \cos \theta A_{in} (\Omega) - e^{i \phi} \sin 
	\theta A_{in} (-\Omega) \nonumber \\
 A_{out1}(-\Omega) &=& i
  \cos \theta v_{in}(-\Omega) - i e^{i \phi} \sin \theta v_{in}
  (-3\Omega)\nonumber \\
   A_{out2}(\Omega) &=& -e^{i \phi} \cos \theta v_{in}(\Omega) - \sin \theta v_{in}(3 \Omega) \nonumber \\
A_{out2} (-\Omega) &=&  i e^{i \phi} \cos \theta 
	A_{in}(-\Omega) + i \sin \theta A_{in} (\Omega)
\end{eqnarray}

\noindent where the absence of hats indicate operators in the Fourier domain and the modulation frequency applied
to the AOM is $2\Omega$.  The diffraction efficiency of the
AOM is $sin^{2} \theta$ and, assuming $100\%$ maximum diffraction efficiency, $\theta$ is proportional to the amplitude
of the radio-frequency signal driving the AOM \cite{young}.  The UMZI is designed to have a differential time delay such that $\omega_0 \tau=\pi/2$ and $ \pm \Omega \tau = \pm \pi/2$ where $\tau = \Delta l/c$ for $\Delta l $ the differential path length and c the speed of light in the UMZI.   The differential time delay between the two paths from the outputs of the UMZI to the AOM is $\tau_2$, such that $\omega_0\tau_2 = \phi$ and $\Omega \tau_2 \approx 0$.   All beamsplitters have power transmission of $50 \%$ and the symmetric phase convention has been used \cite{walls,resch}.

Eqs. \ref{out} are the key theoretical results of this paper.   To illustrate with an explicit example, a single photon state $\mid \psi \rangle= \left( \mu \mid 1 \rangle_{-\Omega} + \nu \mid 1 \rangle_{+\Omega} \right) \otimes \mid 0 \rangle$ would be transformed to $\mid \psi \rangle^{\prime}= \left[ \left( i \mu e^{i \phi} \cos \theta + i \nu \sin \theta \right) \mid 1 \rangle_{-\Omega,A_{out2}}\right.$  $\left.+\left(  \nu\cos \theta - \mu e^{i\phi} \sin \theta \right)  \mid 1 \rangle_{+\Omega,A_{out1}} \right] \otimes \mid 0 \rangle$.  If one or both output ports of the device illustrated in Fig. \ref{schem} are monitored, the input state may be analysed as a function of both the diffraction efficiency and the optical phase at the AOM.   The two degrees of freedom, $\theta$ and $\phi$, allow complete analysis of both the amplitude and relative phase of the projections of the input state onto the two bases.

%\section{Experiment}

\noindent {\it Experiment}: The rf-analyser has been constructed and tested experimentally using a
commercial Nd:YAG laser operating at $1064nm$.  Phase modulation (PM) sidebands
were imposed on the laser beam at $90.5MHz$ with a commercial electro-optic modulator (EOM).  In the absence of
modulation, the output of the laser was quantum noise limited in both
amplitude and phase quadratures at this frequency.  The differential path length in the UMZI was matched to the modulation frequency of $90.5MHz$ at $0.83m$ and the fringe visibility was $98\%$.  The fringe visibility at the AOM was $88\%$.

 Operation was tested using spectral measurements of the $A_{out1}$ beam and a relatively strong modulation
depth -  approximately $1.5\%$ of the original laser power was
transferred to each of the PM sidebands.    Spectral measurements were performed using a scanning confocal Fabry-Perot cavity \cite{kogelnik} with a Free-Spectral Range of
$500 MHz$ and a linewidth of approximately $2 MHz$.  This optical spectrum analyser (OSA) design allowed unambiguous identification of the carrier, frequency shifted carrier and sideband of interest  ($+\Omega$ for the $A_{out1}$ beam).  
The rf-analyser was also tested using weak modulation and Quantum Noise Limited (QNL) homodyne detection (with efficiency of $95\%$) of the $A_{out1}$ beam .  Homodyne detection allows measurements of quadrature amplitude and phase fluctuations - defined respectively in Fourier space as  $X^{+}(\Omega) = A(\Omega) + A(-\Omega)^{\dagger}$  and $X^{-}(\Omega) = i(A(\Omega) - A(-\Omega)^{\dagger})$ \cite{bachor}.  Variances are $V^{\pm} (\Omega) = \langle |X^{\pm}(\Omega)|^{2}
  \rangle$.  We model the input states as independent vacua at $\pm \Omega$ with coherent states superimposed at each of $+\Omega$ and $-\Omega$.  The relative amplitudes and phases of coherent states imposed at $\pm \Omega$ are determined by the modulation scheme used.   The $A_{out2}$ beam was used to
lock to an appropriate relative optical phase of $\omega_0 \tau=\pi/2$ in the UMZI and the desired $\phi$ at the input to the AOM.

The power of the positive sideband of the $A_{out1}$ output (let us denote it $P_{+\Omega, out1}$) can be determined directly from Eq. \ref{out}.  The variance of the output of the rf-analyser at $\Omega$ may be determined using Eq. \ref{out}, the definitions of $X^\pm$ and $V^\pm$ and the relation that $\hat{{A}}(t) 
\rightarrow A(\Omega) \Rightarrow  \hat{{A}}(t)^{\dagger} \rightarrow A(-\Omega)^{\dagger}$  \cite{glauber}.  The ensemble average photon number in a given sideband may be determined from quadrature measurements as $n=(V^++V^--2)/4$ \cite{ralph00}.  For a PM input, we find that:

\begin{equation} \label{PM}
\frac{P_{+\Omega, out1}(\theta,\phi)}{P_{ \Omega, in}} = \frac{ n_{+\Omega, out1}} {n_{ \Omega, in}}= \frac{1}{2} \left( 1 + \sin 2\theta \cos \phi \right) \\
\end{equation}

\noindent where $P_{\Omega, in} = P_{+\Omega, in}+P_{-\Omega, in}=2P_{+\Omega, in}$ for $P_{\pm \Omega, in}$ the input power of the $\pm \Omega$ modulation sidebands respectively.  Similar notational definitions apply for the ensemble average photon numbers.  All input variances except $V_{in}^-$ are set to the QNL. 

      \begin{figure}[!htb]
\begin{center}
  \includegraphics[width=8.3cm]{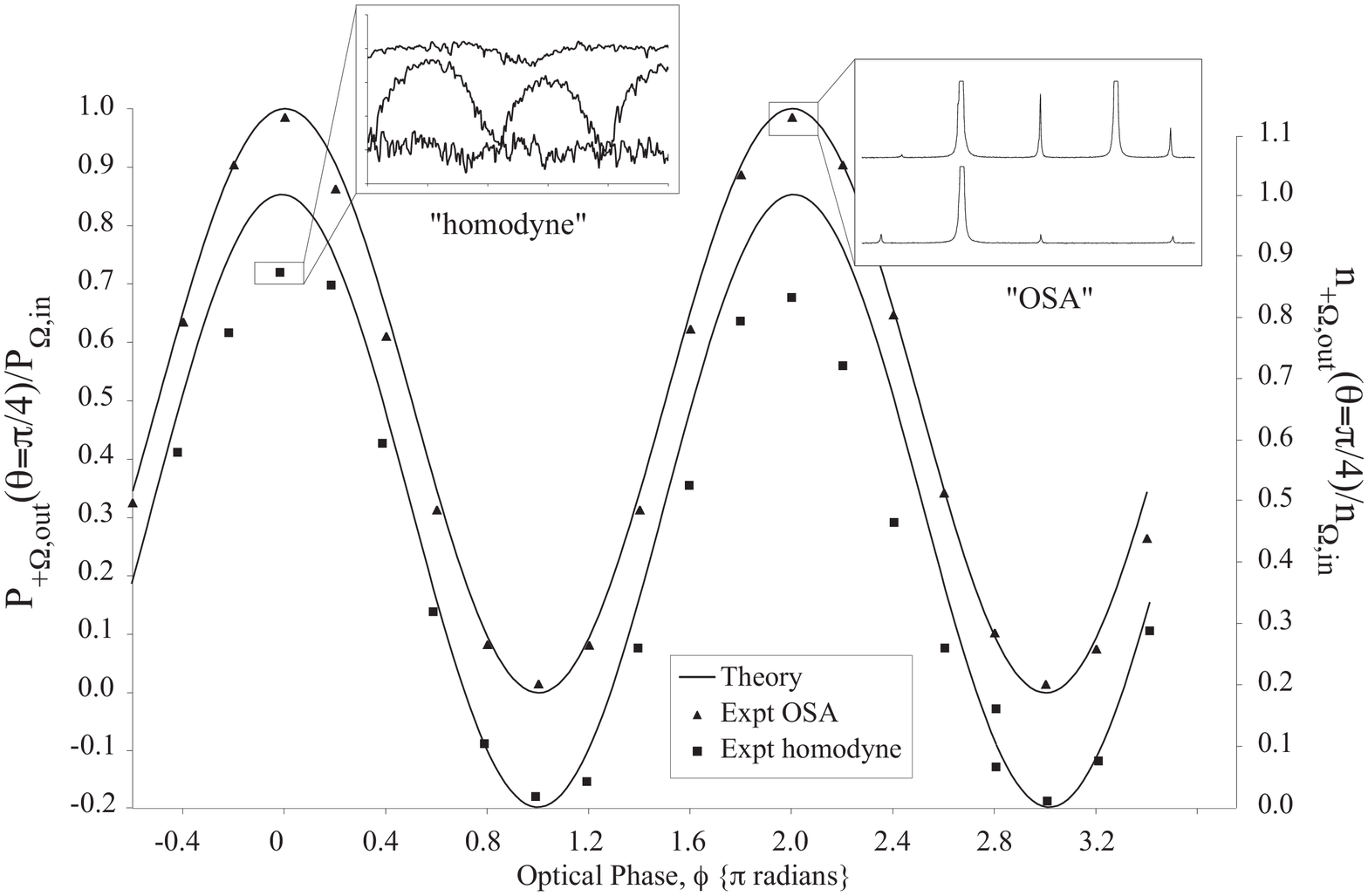}
  \caption
{\label{both_phi} Measurements of $P_{+\Omega,out1}/P_{\Omega,in}$ and $n_{+\Omega,out1} /n_{\Omega,in}$ as a function of the optical phase $\phi$. Also shown are theoretical
predictions for these measurements. The error bars for the spectral measurements are approximately $0.005$ and $12\%$ of the nominal value for the homodyne measurements. }
\end{center}
   \end{figure}

First let us consider an analogy to polarisation analysis in the circular basis.  In this situation, the rf-analyser was operated with fixed $\theta=\pi/4$ and $\phi$ was varied.  Fig. \ref{both_phi} shows measurements of $P_{+\Omega,out1} /P_{\Omega,in}$ and $n_{+\Omega,out1} /n_{\Omega,in}$ for the PM input.  The theoretical predictions for these measurements are also shown.   Both experimental results show the desired sinusoidal behaviour with $\phi$.  The visibility of the fringe based on spectral measurements is $97\%$ and the visibility of the homodyne fringe is $96\%$.    All the results based on homodyne detection are slightly lower than predicted by theory.  This is due to the finite modematching efficiencies in the experiment to which the homodyne detection system was particularly sensitive.  The insets to Fig. \ref{both_phi} show examples of how these measurements were made.  In the spectral measurements, the horizontal axis of the inset represents frequency and the vertical axis shows the optical power transmitted through the scanning OSA.  The lower trace shows a measurement of the attenuated (by a factor of 4) PM input to the rf-analyser.  From the left the peaks are: the lower PM sideband; the truncated carrier; the upper PM sideband; and the mode-mismatch peak for the cavity.  The upper trace (vertically offset for clarity) shows a measurement of the output of the rf-analyser for $[\theta,\phi]=[\pi/4,0]$.  From the left the peaks are: mode-mismatch for the upshifted carrier; the truncated carrier; the sideband of interest; the upshifted carrier; and mode-mismatch for the carrier.  The horizontal axis of the inset for the homodyne measurements is the phase of the homodyne local oscillator  and the vertical axis is the variance of the measurement in dB relative to the QNL.  From bottom to top the traces are: the QNL calibration;  the (attenuated by a factor of 4) PM input to the rf-analyser; and the output of the rf-analyser.  The output of the rf-analyser is single sideband modulation and hence it is phase insensitive (i.e. $V_{out1}^+=V_{out1}^-=V_{out1}$).  The phase quadrature variance of the input was inferred from the measurement of the attenuated input in the standard way \cite{bachor}.  That is,  $V_{in}^-=4V_{in,det}^- - 3$ where $V_{in,det}^-$ is the measured phase quadrature variance.
  \begin{figure}[!htb]
\begin{center}
  \includegraphics[width=8.3cm]{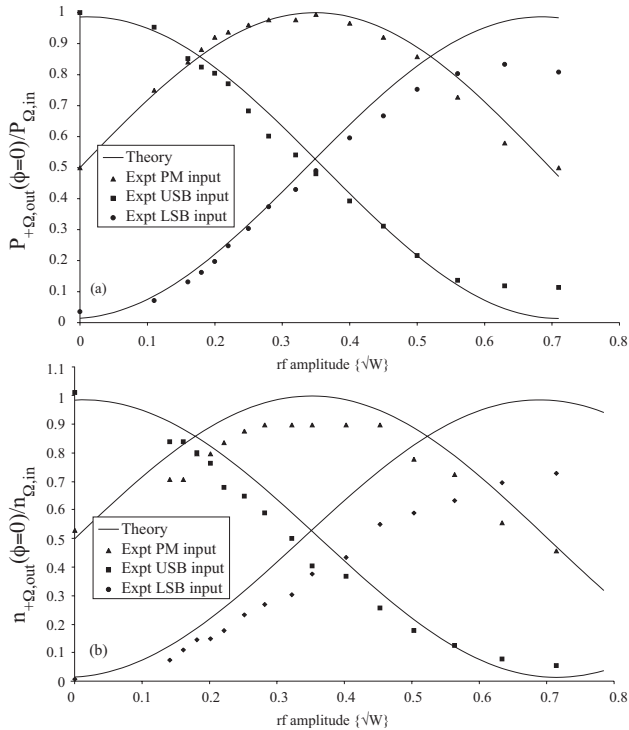}
  \caption{\label{both_theta} Measurements of $P_{+\Omega,out1}/P_{\Omega,in}$ and $n_{+\Omega,out1} /n_{\Omega,in}$ as a function of the drive power to the AOM. Also shown are theoretical predictions for these measurements. The diffraction efficiency of the AOM was measured as a function of the amplitude of the rf drive signal to calibrate the horizontal axis to $\theta$ for the theory curves.  The error bars are as for Fig. \ref{both_theta}. }
\end{center}
   \end{figure}

A true rf-analyser should be able to distinguish between different input states.  Two additional modulation states were used as inputs to test this - lower (upper) sideband modulation, LSB (USB) comprising, nearly, a single modulation sideband
at $-\Omega$ ($+\Omega$).  Both the LSB and USB inputs were derived experimentally by transmitting a PM field through an additional "state preparation" UMZI prior to the rf-analyser.  Our state preparation UMZI had $\Omega \tau \approx 1.33$ radians and was locked to $\omega_0 \tau= \pm \pi/2$.  An UMZI with this value of $\Omega \tau$ would generate $P_{-\Omega,in}=0.985 P_{\Omega,in}$ and $P_{+\Omega,in}=0.015 P_{\Omega,in}$ for the LSB input  and vice versa for the USB input \cite{hunt04}.   The two sidebands would have a relative phase of approximately $1.33$ radians.   

Fig. \ref{both_theta} shows measurements of $P_{+\Omega,out1} /P_{\Omega,in}$ and $n_{+\Omega,out1} /n_{\Omega,in}$ for the PM, LSB and USB input states as a function of the drive power to the AOM for a fixed $\phi=0$.   This measurement is analogous to polarisation analysis in the linear basis.  The theoretical predictions for these measurements are shown as solid lines.    Fig. \ref{both_theta} shows that all three input states may be clearly distinguished on the basis of these measurements.   The results in Fig. \ref{both_theta} show good agreement between theory and experiment up to an AOM drive power of approximately $0.25W$ (i.e. $(0.5 \sqrt{W})^2$).  Beyond this, the experimental results and theoretical predictions diverge from one another.  This is primarily because the AOM had a measured maximum diffraction efficiency of $85\%$ whereas Eq.  \ref{out} assumes $100\%$ maximum diffraction efficiency.   
   
 %\section{Conclusion}
  \noindent {\it Conclusion}: In summary, we have proposed and demonstrated a device which may be
  used to perform arbitrary rotations of states in optical sideband
  modes and to then coherently separate photons from different
  sideband modes into separate spatial modes.  This device is the 
  ``frequency basis'' analogue of a Babinet compensator followed by a 
  polarising beamsplitter.  Such a device may be put to all of the 
  same uses as it's polarisation analogue in both discrete and 
  continuous variable quantum optical experiments.
  
 %\acknowledgements
 \noindent {\it Acknowledgements}: This work was supported by the Australian Research Council.  We  thank James Webb for technical assistance.

\end{document}